\documentclass[preprint,12pt]{elsarticle}

\usepackage{amssymb}
\usepackage{amsmath}
\usepackage{tabularx}
\usepackage{array}
\usepackage{multirow}
\usepackage{booktabs}
\usepackage{empheq}
\usepackage{hyperref}
\usepackage{lineno}
\journal{}

\begin{document}

\begin{frontmatter}

\title{Learning the Exact Flux: Neural Riemann Solvers with Hard Constraints}

%% Author name
\author[label1]{Yucheng Zhang\corref{cor1}}
\author[label1]{Chayanon Wichitrnithed}
\author[label1]{Shukai Cai}
\author[label1]{Sourav Dutta}
\author[label2]{Kyle Mandli}
\author[label1,label3]{Clint Dawson}

%% Corresponding author footnote
\cortext[cor1]{Corresponding author. Email: yucheng@utexas.edu}

\affiliation[label1]{organization={Oden Institute for Computational Engineering and Sciences, University of Texas at Austin},
            city={Austin},
            state={Texas},
            country={USA}}

\affiliation[label2]{organization={National Center for Atmospheric Research},
            city={Boulder},
            state={Colorado},
            country={USA}}

\affiliation[label3]{organization={Department of Aerospace Engineering and Engineering Mechanics, University of Texas at Austin},
            city={Austin},
            state={Texas},
            country={USA}}

\begin{abstract}

\! Godunov-type methods, which obtain numerical fluxes through local Riemann problems at cell interfaces, are among the most fundamental and widely used numerical methods in computational fluid dynamics. Exact Riemann solvers faithfully solve the underlying equations, but can be computationally expensive due to the iterative root-finding procedures they often require. Consequently, most practical computations rely on classical approximate Riemann solvers, such as Rusanov and Roe, which trade accuracy for computational speed. 

Neural networks have recently shown promise as an alternative for approximating exact Riemann solvers, but most existing approaches are data-driven or impose weak constraints. This may result in problems with maintaining balanced states, symmetry breaking, and conservation errors when integrated into a Godunov-type scheme. To address these issues, we propose a hard-constrained neural Riemann solver (HCNRS) and enforce five constraints: positivity, consistency, mirror symmetry, Galilean invariance, and scaling invariance. 

Numerical experiments are carried out for the shallow water and ideal-gas Euler equations on standard benchmark problems. In the absence of hard constraints, violations of the well-balanced property, mass conservation, and symmetry are observed. Notably, in the Euler implosion problem, the exact Riemann solver with MUSCL–Hancock captures the jet structure well, whereas the Rusanov flux is too diffusive and smears it out. HCNRS accurately reproduces the solution obtained by the exact Riemann solver. In contrast, an unconstrained neural formulation lacks mirror symmetry, which makes the solution depend on the choice of flux normal direction. As a result, the jet is either shifted or lost, along with diagonal symmetry.

\end{abstract}


\end{frontmatter}

\section{Introduction}
\label{sec1}

In recent years, neural networks have emerged as a promising approach for accelerating computational fluid dynamics (CFD) simulations. One of the most prevalent methodologies frames fluid simulation as a supervised regression task, where a neural network is trained to map the known state of the flow field to the solution at a subsequent discrete time step. During inference, these models are deployed autoregressively to evolve the system over extended temporal horizons \cite{tompson2017accelerating, brunton2020machine,panchigar2022MLCFD,drikakis2023can}.
Neural ordinary differential equations (Neural ODEs) \cite{chen2018neural} seek to approximate the time derivatives of the governing equations using neural networks, with the network’s effective depth interpreted as a continuous-time integration process. Parallel to these data-driven approaches, physics-informed neural networks (PINNs) \cite{raissi2019physics, zhao2024comprehensive} incorporate the governing equations directly into the loss function, enabling training without labeled solution data by enforcing the PDE residual at collocation points. Closely related developments in neural operator learning aim to approximate solution operators acting between function spaces, providing a mesh-independent mapping from inputs such as initial or boundary conditions to full solution fields. Representative architectures include the Deep Operator Network (DeepONet) \cite{lu2021learning} and the Fourier Neural Operator (FNO) \cite{li2020fourier}.

Despite their promise, neural network based approaches for CFD face challenges when applied to advection-dominated systems. In practice, most existing studies adopt a global surrogate formulation in which the neural network acts on the full spatial domain. While such global representations are better suited for smooth flows, they are more difficult to reconcile with the conservation laws, whose solutions may contain discontinuities. Universal approximation theorems establish convergence in integral norms, offering no theoretical guarantees for pointwise accuracy in the vicinity of these discontinuities. An additional challenge associated with global surrogate models is generalization during inference, particularly in scenarios involving strong variability in external forcing. For example, in storm surge simulations, variations in hurricane track, intensity, and timing can lead to very different flow responses. Accurately capturing such variability within a single global surrogate typically requires extensive training data spanning a high-dimensional parameter space, making training costly and generalization to unseen scenarios difficult.

Rather than constructing a global surrogate, an alternative strategy is to examine the structure of classical numerical methods and target their most computationally intensive components. In Godunov-type schemes, the repeated solution of local Riemann problems at cell interfaces constitutes the most dominant computational costs \cite{toro2013riemann}. Exact Riemann solvers (RS) require the iterative solution of nonlinear algebraic systems and are therefore rarely used in practical large scale simulations. This expense has motivated the widespread adoption of approximate RS, which trade accuracy for efficiency. All of these considerations make the exact RS a natural candidate for surrogate modeling. This approach is less ambitious than constructing a global surrogate, but it addresses several of the challenges discussed above. The underlying mapping defined by the exact Riemann problem and learned by the neural network is continuous, a property that follows from the implicit function theorem. As a result, standard universal approximation theorem guarantees that this mapping can be approximated arbitrarily well by neural networks. Generalization during inference is also greatly improved, as the surrogate needs only approximate a local mapping between left and right states, rather than a high-dimensional global solution operator. Consequently, training can be performed using randomly sampled admissible states without requiring coverage of global flow configurations. Moreover, by only targeting a well-defined component of a classical numerical solver, the resulting hybrid model is inherently less black-box in nature and more closely aligned with established numerical analysis principles.

Early work on neural network based RS was presented by Gyrya et al.\ \cite{gyrya2024machine}, who considered the Euler equations for ideal gas and demonstrated that the neural network predicted solutions can be practically indistinguishable from exact Riemann solutions, provided the training dataset contains sufficient samples with the same wave patterns as the target problem. Magiera et al.\ \cite{magiera2020constraint} proposed constraint-aware neural networks for Riemann problems, enforcing wave speeds derived from the Rankine--Hugoniot conditions either strongly or weakly through two different methods, and integrated the resulting models into a front-capturing numerical scheme. Wang et al.\ \cite{wang2022riemann} and Ruggeri et al.\ \cite{ruggeri2022neural} extended learning-based RS to non-ideal equations of state. In particular, Ruggeri et al.\ provided detailed performance benchmarks across different wave types and reported speedups of more than three orders of magnitude compared to exact RS. Peyvan et al.\ \cite{peyvan2024riemannonets} proposed RiemannONets, a DeepONet-based framework trained using a two-stage procedure following recent work in operator learning. The two-stage procedure results in significantly improved accuracy and enables highly accurate Riemann solutions suitable for real-time forecasting. Wu et al.\ \cite{wu5295012rimnet} focused on the shallow water equations and proposed learning-based approximations of the Harten, Lax and van Leer (HLL) Riemann solver. Their results showed low relative $L^1$ and $L^2$ errors, along with computational speedups ranging from approximately $15\%$ to $30\%$.

While the studies reviewed above demonstrate the promise of learning based RS, most existing approaches rely primarily on data-driven training and weak regularization. When neural RS without hard constraints are embedded into numerical schemes, several issues may arise. One important issue is the loss of the well-balanced property. Ruggeri et al.\ \cite{ruggeri2022neural} noted that unconstrained neural network based solvers introduce spurious oscillations for constant initial conditions. Similarly, Wu et al.\ \cite{wu5295012rimnet} demonstrated that violations of the well-balanced property persist even when the training dataset is enriched with still water states and the still water condition is weakly enforced, although such measures can partially reduce the error. A second issue concerns the enforcement of solid-wall boundary conditions. In the absence of hard constraints, neural RS produce nonzero numerical fluxes at solid walls for quantities such as water depth in the shallow water equations or density in the Euler equations. This behavior leads to a failure of mass conservation for the numerical scheme.

Motivated by these issues, we propose a neural RS that embeds key constraints strongly into the solver architecture. Specifically, we identify five fundamental constraints, positivity, consistency, mirror symmetry, Galilean invariance, and scaling invariance, and enforce them strongly by construction. Among these, consistency ensures preservation of well-balanced properties, while mirror symmetry guarantees correct solid-wall boundary behavior. Galilean invariance and scaling invariance reduce the complexity of the learning problem by reducing the input dimension.

The remainder of this paper is organized as follows. Section 2 looks at the role of the Riemann problem in Godunov-type schemes. It also summarizes important features of the exact RS for the shallow water and compressible Euler equations for ideal gas, which are relevant to this study. Section 3 presents the proposed constraint framework and the neural solver design. Section 4 shows numerical results for the shallow water and Euler equations, including training settings. Section 5 discusses computational cost. Section 6 concludes with a summary of the method and experiments with considerations for future directions.

\section{Riemann Solvers in Godunov-type schemes}
We consider a system of hyperbolic conservation laws written in one spatial dimension as
\begin{equation}
\partial_t \mathbf{U} + \partial_x \mathbf{F}(\mathbf{U}) = \mathbf{0},
\label{eq:conservation_law}
\end{equation}
where $\mathbf{U}(x,t) \in \mathbb{R}^m$ denotes the vector of conservative variables and
$\mathbf{F}(\mathbf{U})$ is the corresponding flux function.

\subsection{Godunov-Type Discretization}
Godunov-type methods approximate Eq.~\eqref{eq:conservation_law}
by evolving discrete states within spatial elements and computing numerical fluxes at their interfaces through the solution of Riemann problems.
The formulation can be written abstractly as
\begin{equation}
\frac{d}{dt} \mathbf{U}_i(t)
= -\frac{1}{\Delta x}
\left(
\mathbf{F}_{i+\frac12} - \mathbf{F}_{i-\frac12}
\right),
\label{eq:godunov_update}
\end{equation}
where $\mathbf{U}_i(t)$ denotes the average over cell $i$, and $\mathbf{F}_{i\pm\frac12}$ are numerical fluxes defined at cell interfaces.

The numerical flux is obtained by solving a local Riemann problem, that is, the initial value problem for Eq.~\eqref{eq:conservation_law} with piecewise-constant initial data
\begin{equation}
\mathbf{U}(x,0) =
\begin{cases}
\mathbf{U}_L, & x \le 0, \\
\mathbf{U}_R, & x > 0,
\end{cases}
\label{eq:riemann_ic}
\end{equation}
where $\mathbf{U}_L$ and $\mathbf{U}_R$ denote the left and right states immediately adjacent to the interface.

\begin{figure}
    \centering
    \includegraphics[width=1.0\linewidth]{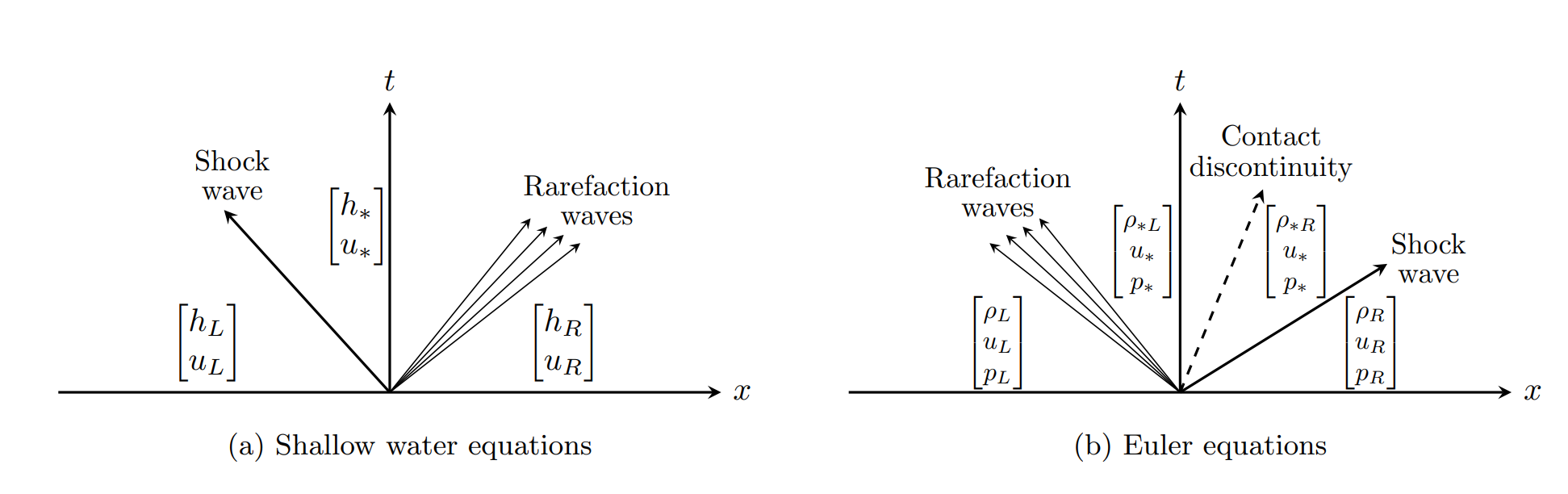}
    \caption{
    Typical solutions of the Riemann problem.
    (a) Shallow water equations: a left-going shock and a right-going rarefaction wave.
    (b) Euler equations: a left-going rarefaction, a contact discontinuity, and a right-going shock. $h$ and $u$ denote water depth and velocity for the shallow water equations, while $\rho$, $u$, and $p$ denote density, velocity, and pressure for the Euler equations. Subscripts $L$ and $R$ denote the left and right initial states, respectively, and the subscript $*$ indicates intermediate (star) states. $*L$ and $*R$ denote left and right star-region states. For the Euler equations, the velocity and pressure are uniform across the contact discontinuity, while the density may differ.
    }
    \label{fig:demo}
\end{figure}

The exact RS computes the solution of the initial value problem \eqref{eq:riemann_ic} for $t>0$ exactly. The solution is self-similar and depends only on $\xi = x/t$. It consists of a sequence of elementary waves separating constant intermediate states, typically denoted by star states ($*$). For the shallow water or Euler equations, the solution structure includes shock waves and rarefaction waves, and, in the case of the Euler equations, an additional contact discontinuity. Fig.~\ref{fig:demo} illustrates typical solution structures. For the shallow water equations, the solution shown consists of a left-going shock and a right-going rarefaction wave, separating the initial states from a single intermediate state. For the Euler equations, the solution consists of a left-going rarefaction wave, a right-going shock, and an intermediate contact discontinuity separating two star states. Across the contact discontinuity, the velocity and pressure remain continuous, while the density may exhibit a jump. The intermediate state is determined by enforcing the Rankine--Hugoniot conditions across shocks and tracing the integral curves for rarefactions.

For both the shallow water and Euler equations, the Riemann problem reduces to a nonlinear scalar equation for the intermediate water height (or pressure), obtained by matching wave curves from the left and right states. This equation is solved using a root-finding procedure, after which the complete wave structure and flux at the interface are uniquely determined.

\subsection{Exact Riemann Solvers for the Shallow Water and Euler Equations}

In this subsection, we briefly state the associated root-finding problem and the resulting intermediate state calculations required for flux evaluation for shallow water and ideal-gas Euler equations. Once the intermediate states are obtained, the solution to the Riemann problem is sampled at the interface ($x/t = 0$) to determine the Godunov interface state, from which the numerical flux is evaluated. For the shallow water equations, we follow the formulation in \cite{toro2001shock}. For the Euler equations, we use the approach described in \cite{toro2013riemann}. More details can also be found in \cite{leveque2002finite, ketcheson2020riemann}.

\subsubsection{Shallow Water Equations}
The one dimension shallow water equations take the form
\begin{equation}
\begin{aligned}
h_t + (hu)_x &= 0, \\
(hu)_t + \left( hu^2 + \tfrac{1}{2} g h^2 \right)_x & = 0,
\end{aligned}
\label{eq:swe}
\end{equation}
where $h$ denotes the water depth, $u$ the depth-averaged velocity, $g$ the gravitational acceleration.

Given left and right states $(h_L,u_L)$ and $(h_R,u_R)$, the solution of the Riemann problem consists of an intermediate star states $(h_*,u_*)$. The root finding problem is defined by
\begin{equation}
\phi(h_*) = f_L(h_*) + f_R(h_*) + (u_R - u_L) = 0,
\label{eq:phi_root}
\end{equation}
where the functions $f_L$ and $f_R$ represent the left and right going wave curves, respectively.

For either side $\alpha \in \{L,R\}$, we define the wave curve associated with the state $(h_\alpha,u_\alpha)$ by
\begin{equation}
f_\alpha(h) =
\begin{cases}
2\left(\sqrt{g h} - \sqrt{g h_\alpha}\right),
& h \le h_\alpha, \\[0.6em]
(h - h_\alpha)\sqrt{\dfrac{g}{2}\left(\dfrac{1}{h} + \dfrac{1}{h_\alpha}\right)},
& h > h_\alpha.
\end{cases}
\label{eq:wave_curve}
\end{equation}

Once the root $h_*$ of \eqref{eq:phi_root} has been found, the star-state velocity $u_*$ can be computed as
\begin{equation}
u_* =
\frac{1}{2}\left(u_L + u_R\right)
+ \frac{1}{2}\left( f_R(h_*) - f_L(h_*) \right),
\label{eq:u_star_symmetric}
\end{equation}

\subsubsection{Euler Equations}
The one-dimensional compressible Euler equations for an ideal gas take the form
\begin{equation}
\begin{aligned}
\rho_t + (\rho u)_x &= 0, \\
(\rho u)_t + \left(\rho u^2 + p\right)_x &= 0, \\
E_t + \left(u(E+p)\right)_x &= 0,
\end{aligned}
\label{eq:euler}
\end{equation}
where $\rho$ denotes the density, $u$ the velocity, $p$ the pressure, and $E$ the total energy density.

The total energy is given by
\[
E = \rho\left(\frac{1}{2}u^2 + e\right),
\]
where $e = e(\rho,p)$ is the specific internal energy given by a caloric Equation of State. For an ideal gas,
\[
e = \frac{p}{(\gamma - 1)\rho},
\]
where $\gamma$ is the ratio of specific heats.

Given left and right states $(\rho_L,u_L,p_L)$ and $(\rho_R,u_R,p_R)$, the exact Riemann solution consists of nonlinear waves separating constant states, with intermediate star states $(\rho_{L*},\rho_{R*},u_*,p_*)$. The star pressure $p_*$ is determined as the solution of the scalar nonlinear equation
\begin{equation}
\phi(p_*) = f_L(p_*) + f_R(p_*) + (u_R - u_L) = 0,
\label{eq:euler_phi_root}
\end{equation}
where $f_L$ and $f_R$ denote the left- and right-going wave curves in pressure
space.

For either side $\alpha \in \{L,R\}$, the wave curve associated with the state
$(\rho_\alpha,u_\alpha,p_\alpha)$ is defined by
\begin{equation}
f_\alpha(p) =
\begin{cases}
\dfrac{2c_\alpha}{\gamma-1}
\left[\left(\dfrac{p}{p_\alpha}\right)^{\frac{\gamma-1}{2\gamma}} - 1\right],
& p \le p_\alpha \\[1.0em]
(p - p_\alpha)\sqrt{\dfrac{2}{(\gamma+1)\rho_\alpha}}
\left(p + \dfrac{\gamma-1}{\gamma+1}p_\alpha\right)^{-\frac{1}{2}},
& p > p_\alpha
\end{cases}
\label{eq:euler_wave_curve}
\end{equation}
where $c_\alpha = \sqrt{\gamma p_\alpha / \rho_\alpha}$ is the sound speed in the
state $\alpha$.

Once the root $p_*$ of \eqref{eq:euler_phi_root} has been found, the star velocity
can be calculated as
\begin{equation}
u_* =
\frac{1}{2}\left(u_L + u_R\right)
+ \frac{1}{2}\left( f_R(p_*) - f_L(p_*) \right),
\label{eq:euler_u_star}
\end{equation}

Once the star pressure $p_*$ and star velocity $u_*$ have been determined, the
star-state density $\rho_{\alpha*}$ can be computed as

\begin{equation}
\rho_{\alpha*} =
\begin{cases}
\rho_\alpha \left( \dfrac{p_*}{p_\alpha} \right)^{\frac{1}{\gamma}},
& p_* \le p_\alpha, \\[1.0em]
\rho_\alpha \,
\dfrac{p_* + \dfrac{\gamma-1}{\gamma+1}p_\alpha}
{\dfrac{\gamma-1}{\gamma+1}p_* + p_\alpha},
& p_* > p_\alpha.
\end{cases}
\label{eq:euler_rho_star}
\end{equation}

\section{Methodology}
In this section, we present a set of hard constraints that we want to enforce in the neural RS. We then state the overall workflow used to incorporate these constraints into the model design.

\subsection{Hard Constraints}
We identify and enforce the following five hard constraints:
\begin{enumerate}
  \item \textbf{Positivity:}  
  The Riemann solver must preserve the non-negativity of physically positive quantities.

  \item \textbf{Consistency:}  
  When the left and right states coincide, the numerical flux must reduce to the flux function evaluated at that state.

  \item \textbf{Mirror Symmetry:}  
  Exchanging the left and right states while reversing the velocity components implies that the flux changes sign accordingly, with its magnitude preserved.
  
  \item \textbf{Galilean invariance:}  
  If a constant velocity is added to both the left and right states, the resulting numerical flux must transform consistently with the Galilean transformation.

  \item \textbf{Scaling invariance:}  
  Under certain scalings of the state variables, both rarefaction and shock curves scale consistently, and the resulting numerical flux must scale accordingly.
\end{enumerate}

To state the constraints in a uniform manner, we introduce the Riemann solution operator $\mathcal{R}$, which maps left and right states to the intermediate star states. For the shallow water equations, $\mathcal{R}_{\mathrm{SWE}}(h_L,u_L,h_R, u_R)=(h_*,u_*)$, whereas for the Euler equations, $\mathcal{R}_{\mathrm{Euler}}(\rho_L,u_L,p_L,\rho_R,u_R,p_R)=(\rho_{L*},\rho_{R*},u_*,p_*)$. The corresponding constraint statements for each system are summarized in Table~\ref{tab:table1}.

\begin{table*}[t]
\caption{Hard physical constraints imposed on the Riemann operators
$\mathcal{R}_{\mathrm{SWE}}$ and $\mathcal{R}_{\mathrm{Euler}}$.}
\label{tab:table1}

% \footnotesize
\scriptsize
\setlength{\tabcolsep}{5pt}
\renewcommand{\arraystretch}{1.22}

\begin{tabularx}{\textwidth}{@{}p{0.16\textwidth} X X@{}}
\toprule
\textbf{Constraint} & \textbf{Shallow Water Equations} & \textbf{Euler Equations} \\
\midrule
Positivity
& $h_*\geq0$
& $\rho_{L*}\geq0,\;\rho_{R*}\geq0,\;p_*\geq0$ \\
\midrule

Consistency
& $\mathcal{R}_{\mathrm{SWE}}(h,u,h,u)=(h,u)$
& $\mathcal{R}_{\mathrm{Euler}}(\rho,u,p,\rho,u,p)=(\rho,u,p)$ \\
\midrule

\multirow{2}{0.16\textwidth}{Mirror \\ Symmetry}
& $\mathcal{R}_{\mathrm{SWE}}(h_R,-u_R,h_L,-u_L)$
& $\mathcal{R}_{\mathrm{Euler}}(\rho_R,-u_R,p_R,\rho_L,-u_L,p_L)$ \\
& $= (h_*,-u_*)$
& $= (\rho_{R*},\rho_{L*},-u_*,p_*)$ \\
\midrule

\multirow{2}{0.16\textwidth}{Galilean \\ Invariance}
& $\mathcal{R}_{\mathrm{SWE}}(h_L,u_L+U,h_R,u_R+U)$
& $\mathcal{R}_{\mathrm{Euler}}(\rho_L,u_L\!+\!U,p_L,\rho_R,u_R\!+\!U,p_R)$ \\
& $= (h_*,u_*+U)$
& $= (\rho_{L*},\rho_{R*},u_*+U,p_*)$ \\
\midrule

\multirow{2}{0.16\textwidth}{Scaling \\ Invariance}
& $\mathcal{R}_{\mathrm{SWE}}(\lambda h_L,\sqrt{\lambda}u_L,\lambda h_R,\sqrt{\lambda}u_R)$
& $\mathcal{R}_{\mathrm{Euler}}(\alpha\rho_L,\! \beta u_L,\! \alpha\beta^2 p_L,\! \alpha\rho_R,\beta u_R,\alpha\beta^2 p_R)$ \\
& $= (\lambda h_*,\sqrt{\lambda}u_*)$
& $= (\alpha\rho_{L*},\alpha\rho_{R*},\beta u_*,\alpha\beta^2 p_*)$ \\
& $\forall\,\lambda>0$
& $\forall\,\alpha>0,\;\beta\neq 0$ \\

\bottomrule
\end{tabularx}
\end{table*}

\subsection{Hard Constrained Neural Riemann Solver Architecture}
In this subsection, we describe how the five constraints introduced above are enforced strongly within the HCNRS.

\subsubsection{Galilean and Scaling Invariance}
The Galilean and scaling invariance are first enforced through invariant-based dimensionality reduction. For the shallow water equations, we consider learning $h_*$ obtained from the root-finding procedure. Assume \begin{equation}
\mathcal{NN}_{\mathrm{SWE}}( h_L,u_L, h_R,u_R)
=h_*
\end{equation}
Galilean and scaling invariance imply that, for any $\lambda>0$ and any constant velocity shift $U$,
\begin{equation}
\mathcal{NN}_{\mathrm{SWE}}(\lambda h_L,\sqrt{\lambda}u_L+U,\;\lambda h_R,\sqrt{\lambda}u_R+U)
=\lambda h_*
\end{equation}

Define 
\[
h_{\mathrm{ref}}:=\frac{h_L+h_R}{2},
\qquad
\Delta u:=u_L-u_R.
\]
Choose $\lambda = 1/h_{\mathrm{ref}}$ and $U=-\sqrt{\lambda}\,u_R$. Then
\begin{equation}
\mathcal{NN}_{\mathrm{SWE}}\!\left(\frac{h_L}{h_{\mathrm{ref}}},\frac{\Delta u}{\sqrt{h_{\mathrm{ref}}}},\frac{h_R}{h_{\mathrm{ref}}},0\right)
= \frac{h_*}{h_{\mathrm{ref}}}.
\end{equation}

Note that the first and third normalized depth inputs are not independent:
\begin{equation}
\frac{h_L}{h_{\mathrm{ref}}}+\frac{h_R}{h_{\mathrm{ref}}}=2.
\end{equation}
Hence the depth dependence can be represented by a single dimensionless contrast parameter. We define
\[
\delta_h:=\frac{h_L-h_R}{h_L+h_R}\in(-1,1),
\]
which uniquely determines $(h_L/h_{\mathrm{ref}},\,h_R/h_{\mathrm{ref}})=(1+\delta_h,\,1-\delta_h)$.
To obtain a dimensionless velocity-jump input, we normalize the input by $1/\sqrt{g}$. Therefore, $h_*$ can be learned as a function of two dimensionless inputs multiplied by $h_{ref}$:
\begin{equation}
h_* = h_{ref}\mathcal{NN}_{\mathrm{SWE}}(\delta_h,\delta_u),
\qquad
\delta_u:=\frac{\Delta u}{\sqrt{g\,h_{\mathrm{ref}}}}
\end{equation}

For the ideal-gas Euler equations, we consider learning the star-state pressure $p_*$ obtained from the root-finding procedure. Assume
\begin{equation}
\mathcal{NN}_{\mathrm{Euler}}(\rho_L,u_L,p_L,\rho_R,u_R,p_R)
= p_* .
\end{equation}

Galilean and scaling invariance imply that, for any $\alpha>0$, $\beta\neq 0$, and any constant velocity shift $U$,
\begin{equation}
\begin{aligned}
\mathcal{NN}_{\mathrm{Euler}}(&\alpha\rho_L,\;\beta u_L + U,\;\alpha\beta^2 p_L, \\
&\alpha\rho_R,\;\beta u_R + U,\;\alpha\beta^2 p_R)
= \alpha\beta^2\, p_* .
\end{aligned}
\end{equation}

Define the reference density and pressure
\[
\rho_{\mathrm{ref}} := \frac{\rho_L+\rho_R}{2},
\qquad
p_{\mathrm{ref}} := \frac{p_L+p_R}{2},
\qquad
\Delta u := u_L - u_R.
\]
Choose
\[
\alpha = \frac{1}{\rho_{\mathrm{ref}}},
\qquad
\beta = \sqrt{\frac{\rho_{\mathrm{ref}}}{p_{\mathrm{ref}}}},
\qquad
U = -\beta u_R .
\]
Then the normalized inputs satisfy
\begin{equation}
\mathcal{NN}_{\mathrm{Euler}}\!\left(
\frac{\rho_L}{\rho_{\mathrm{ref}}},
\frac{\Delta u}{\sqrt{p_{\mathrm{ref}}/\rho_{\mathrm{ref}}}},
\frac{p_L}{p_{\mathrm{ref}}},
\frac{\rho_R}{\rho_{\mathrm{ref}}},
0,
\frac{p_R}{p_{\mathrm{ref}}}
\right)
=
\frac{p_*}{p_{\mathrm{ref}}}.
\end{equation}

Similar to the shallow water equations case, for the Euler equations note that the normalized density and pressure inputs are not independent:

\begin{equation}
\frac{\rho_L}{\rho_{\mathrm{ref}}}
+
\frac{\rho_R}{\rho_{\mathrm{ref}}}
= 2,
\qquad
\frac{p_L}{p_{\mathrm{ref}}}
+
\frac{p_R}{p_{\mathrm{ref}}}
= 2.
\end{equation}
Hence, the dependence can be represented by dimensionless contrast parameters
\[
\delta_\rho :=
\frac{\rho_L-\rho_R}{\rho_L+\rho_R},
\qquad
\delta_p :=
\frac{p_L-p_R}{p_L+p_R},
\]
which uniquely determine the normalized pairs. Defining
\[
\delta_u :=
\frac{\Delta u}
{\sqrt{p_{\mathrm{ref}}/\rho_{\mathrm{ref}}}},
\]
the star pressure admits the invariant representation
\begin{equation}
p_* =
p_{\mathrm{ref}}\,
\mathcal{NN}_{\mathrm{Euler}}
(\delta_\rho,\delta_p,\delta_u).
\end{equation}

\subsubsection{Mirror Symmetry, Consistency and Positivity}

After dimension reduction, mirror symmetry changes to
\begin{align}
\mathcal{NN}_{\mathrm{SWE}}(-\delta_h,\delta_u)
&=
\mathcal{NN}_{\mathrm{SWE}}(\delta_h,\delta_u), \\
\mathcal{NN}_{\mathrm{Euler}}(-\delta_\rho,-\delta_p,\delta_u)
&=
\mathcal{NN}_{\mathrm{Euler}}(\delta_\rho,\delta_p,\delta_u)
\end{align}
and the consistency constraint becomes
\begin{align}
\mathcal{NN}_{\mathrm{SWE}}(0,0) &= 1, \\
\mathcal{NN}_{\mathrm{Euler}}(0,0,0) &= 1.
\end{align}

Mirror symmetry requires 
\begin{equation}
\mathcal{NN}(\mathcal{T}\bold{u}) = \mathcal{NN}(\bold{u}),
\end{equation}
where $\mathcal{T}$ denotes the transformation induced by
mirror symmetry, and $I$ is the identity operator.
For the shallow water and Euler cases, $\mathcal{T}$ acts as
\begin{align}
\mathcal{T}_{\mathrm{SWE}}(\delta_h, \delta_u)
&= (-\delta_h, \delta_u), \\
\mathcal{T}_{\mathrm{Euler}}(\delta_\rho, \delta_p, \delta_u)
&= (-\delta_\rho, -\delta_p, \delta_u),
\end{align}

Since $\mathcal{T}^2 = I$, mirror symmetry can be enforced
by symmetrizing an arbitrary function $f$,
\begin{equation}
\mathcal{NN}(\mathbf{u})
=f(\mathbf{u}) + f(\mathcal{T}\mathbf{u})
,
\end{equation}
Consistency is imposed by subtracting the value at the origin and
adding 1,
\begin{equation}
\mathcal{NN}(\mathbf{u})
=
1 + \Big(
f(\mathbf{u}) + f(\mathcal{T}\mathbf{u})
- 2f(\mathbf{0})
\Big),
\end{equation}
so that $\mathcal{NN}(\mathbf{0}) = 1$.  We approximate $f$ using a multilayer perceptron (MLP), and therefore write $f = \mathrm{MLP}$.

Positivity of the star states can be enforced through a simple post-processing step that clips the predicted values to remain strictly positive. However, it requires special careful handling in numerical schemes where positivity is critical, such as in the presence of wet/dry fronts \cite{bunya2009wetting, castro2005numerical}. We state that dry and vacuum states are not the focus of this work, and no such states were encountered in the benchmark problems considered.

For clarity, we restate the star states approximations below:

\noindent\fbox{%
\begin{minipage}{0.97\linewidth}
\textbf{Shallow Water Equations}
\[
\begin{aligned}
h_* &\approx
h_{\mathrm{ref}}
\left(
1 +
\Big(
\mathrm{MLP}(\delta_h,\delta_u)
+
\mathrm{MLP}(-\delta_h,\delta_u)
-
2\,\mathrm{MLP}(\mathbf{0})
\Big)
\right),
\\[2mm]
h_{\mathrm{ref}} &:= \frac{h_L + h_R}{2},\qquad
\delta_h := \frac{h_L - h_R}{h_L + h_R},\qquad
\delta_u := \frac{u_L - u_R}{\sqrt{g\,h_{\mathrm{ref}}}}.
\end{aligned}
\]
\end{minipage}%
}

\noindent\fbox{%
\begin{minipage}{0.97\linewidth}
\textbf{Euler Equations}
\[
\begin{aligned}
p_* &\approx
p_{\mathrm{ref}}
\left(
1 +
\Big(
\mathrm{MLP}(\delta_\rho,\delta_p,\delta_u)
+
\mathrm{MLP}(-\delta_\rho,-\delta_p,\delta_u)
-
2\,\mathrm{MLP}(\mathbf{0})
\Big)
\right),
\\[2mm]
p_{\mathrm{ref}} &:= \frac{p_L + p_R}{2}, \qquad
\rho_{\mathrm{ref}} := \frac{\rho_L + \rho_R}{2},
\\[2mm]
\delta_\rho &:= \frac{\rho_L - \rho_R}{\rho_L + \rho_R}, \qquad
\delta_p := \frac{p_L - p_R}{p_L + p_R}, \qquad
\delta_u := \frac{u_L - u_R}{c_{\mathrm{ref}}},
\\[2mm]
c_{\mathrm{ref}} &:= \sqrt{\frac{p_{\mathrm{ref}}}{\rho_{\mathrm{ref}}}}.
\end{aligned}
\]
\end{minipage}%
}

\section{Numerical Results}
This section evaluates the performance of the proposed method through several numerical benchmarks. We first describe how the training data is generated, the training process, and the settings for the hyperparameters. We then examine the still water and radial dam break test cases for the shallow water equations, as well as the implosion problem for the Euler equations, using various RS. The Rusanov flux, also known as the local Lax-Friedrichs flux, is a commonly used approximate RS and serves as a baseline in this work, it is summarized in~\ref{app:rusanov}. Depending on context, labels such as Exact RS, HCNRS, and Rusanov refer to the full numerical scheme in which the corresponding RS is integrated, or to the RS itself. 

\subsection{Training Settings}

Unlike many global surrogate approaches that require full CFD simulation data, 
training samples here are generated directly by querying the exact RS. For most target applications, two observations allow us to restrict sampling to a substantially smaller and more relevant subset of the full state space:

\begin{enumerate}
  \item \textbf{Bounded variable range.}
  The variables encountered in the target simulation are known \emph{a priori} to lie within an interval.
  
  \item \textbf{Similarity of left and right states.}
  In typical CFD simulations, the left and right states passed into the RS at most interfaces are close, except near strong discontinuities. Therefore, a large fraction of Riemann problems arise from small perturbations around a common background state.
\end{enumerate}

Based on these structural properties, we construct the training set by combining two sampling strategies. First, we draw independent left/right states uniformly within the prescribed interval to have global coverage. Second, we draw single states and form nearby pairs by adding random perturbations. Once trained, the neural RS is intended to work on any simulations whose states remain within the prescribed interval. 

In practical applications, if the admissible variable ranges are not known \emph{a priori}, they may be estimated by running representative simulations for the target application (using either exact or approximate RS) and extending the observed ranges with a suitable tolerance. Training data can also be generated by running full numerical simulations and extracting left and right states at cell interfaces. In addition, online training may be employed to mitigate out-of-range inputs. More optimal or adaptive sampling strategies will be addressed in future work.

We now apply the proposed sampling strategy to the shallow water equation benchmarks. We train a single HCNRS model that is used for both the still water and radial dam break test cases. First, we randomly sample $10{,}000{,}000$ independent left and right states $(h_L,u_L)$ and $(h_R,u_R)$ from the ranges $h \in [0.1, 12]$,  $u \in [-6, 6]$ to ensure global coverage of the admissible state space. Second, we generate $50{,}000{,}000$ close state pairs. We select background states from the same ranges and then construct the left and right states by adding uniformly distributed perturbations in the range $[-5\%, 5\%]$ relative to the background values. Dry states are filtered out using the criterion: a middle state is dry if and only if $u_L + 2\sqrt{g h_L} < u_R - 2\sqrt{g h_R}.$

For the Euler equation benchmarks, training states are sampled from
$
\rho \in [0.1, 4.5] 
$, $
u \in [-2, 2]
$, $
p \in [0.1, 9.5].
$
We again use $10{,}000{,}000$ purely random state pairs and $50{,}000{,}000$ perturbed pairs with $5\%$ relative noise. Vacuum states are excluded using the vacuum condition (see Eq.~(4.82) in \cite{toro2013riemann}).

The root-finding problems are solved using the Newton–Raphson method with a tolerance of $10^{-9}$, initialized with the two-rarefaction approximation. Although $60{,}000{,}000$ samples may appear large, this is comparable to only about 120 time steps (forward Euler) on a $500 \times 500$ computational grid. Data generation is completed within 20 minutes for both equations. The dataset is split into $90\%$ for training and $10\%$ for validation.

For both the shallow water and Euler equations, the HCNRS employs an MLP with three hidden layers of widths $(32,32,32)$ and $\tanh$ activation functions. We also evaluated alternative architectures, including a shallower network $(32,32)$ and deeper or wider variants such as $(32,32,32,32)$ and $(64,32,32)$. The $(32,32)$ configuration resulted in reduced accuracy, whereas the deeper and wider variants achieved similar or slightly worse validation loss while having a larger number of trainable parameters.

For comparison, we also consider a commonly used unconstrained learning baseline in
which star-region states are predicted directly from the left and right
input states using an MLP. This model is referred to as the unconstrained neural Riemann solver (UCNRS). For UCNRS, we use the same hidden-layer widths for consistency. We apply min–max normalization to rescale the data to $[-1,1]$.

All networks are implemented in JAX \cite{jax2018github}. Training is performed using the Adam optimizer \cite{kingma2014adam} with a learning rate $\eta=10^{-3}$. The training process for both models is computationally inexpensive and requires less than 30 minutes for all cases.

\subsection{Still Water}
For still water, UCNRS fails to preserve the well-balanced property. More egregiously, it produces nonzero mass (water depth) fluxes at solid wall boundaries, making the scheme non-conservative. In contrast, HCNRS resolves both issues. The consistency requirement maintains the well-balanced property, while the mirror symmetry guarantees zero numerical fluxes at wall boundaries.

We use the first order well-balanced scheme proposed by Audusse et al.\ \cite{audusse2004fast}. This method relies on a local hydrostatic reconstruction and ensures well-balancedness as long as the consistency condition for the numerical flux is met.

The one dimension shallow water equations with bathymetry take the form
\begin{equation}
\begin{aligned}
h_t + (hu)_x &= 0, \\
(hu)_t + \left( hu^2 + \tfrac{1}{2} g h^2 \right)_x & = -g h z_x,
\end{aligned}
\label{eq:swe}
\end{equation}
where $h$ denotes the water depth, $u$ the depth-averaged velocity, $g = 9.81$ the gravitational acceleration and z(x) the varying topography.

We consider a one-dimensional domain $x \in [0,1]$ discretized using $N=400$
uniform cells. The bathymetry $z(x)$ consists of a smooth localized bump centered at $x=0.5$ and
supported on the interval $x \in [0.3,\,0.7]$, defined as
\[
z(x) =
\begin{cases}
5 \cos^2\!\left( \dfrac{\pi (x - 0.5)}{0.4} \right), & |x - 0.5| \le 0.2, \\
0, & \text{otherwise},
\end{cases}
\]

A lake-at-rest equilibrium is initialized with the water depth $h=10$ outside the support of the bump and decreases to $h=5$ at $x=0.5$. Solid wall boundary conditions are imposed at both ends of the domain. Time stepping is performed using a forward Euler scheme with time step $dt = 10^{-4}$ for 1000 steps, corresponding to a final time $T = 0.1$ and $\mathrm{CFL} \approx 0.4$.

\begin{figure}
    \centering
    \includegraphics[width=0.7\linewidth]{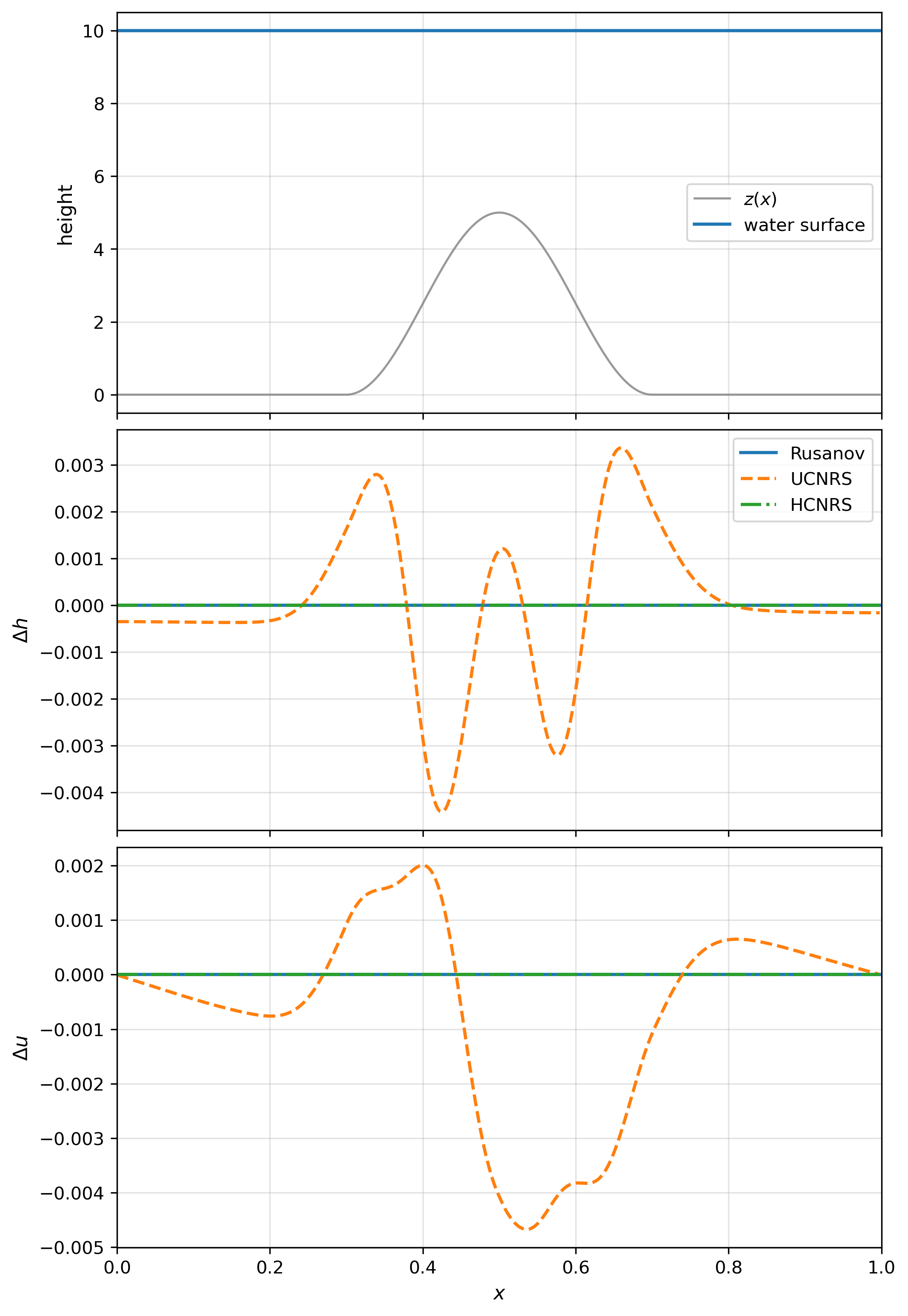}
    \caption{
    Still water test at $t = 0.1\,\mathrm{s}$.
    Top: initial condition showing bathymetry and water surface elevation.
    Middle: deviation in water depth $\Delta h = h - h_{\mathrm{ref}}$, 
    where $h_{\mathrm{ref}}$ denotes the equilibrium depth.
    Bottom: deviation in velocity $\Delta u = u - u_{\mathrm{ref}}$, 
    with $u_{\mathrm{ref}} = 0$.
    }
    \label{fig:well-balanced}
\end{figure}

Fig.~\ref{fig:well-balanced} shows the initial condition, predicted water surface elevation, and velocity profiles at $t = 0.1$. It compares HCNRS, UCNRS, and the Rusanov scheme. The exact RS is omitted from the figure, as it produces identical results to HCNRS. In this still water case, the left and right states passed into both RS are equal ($h_L = h_R$, $u_L = u_R$) at all interfaces (besides the wall boundaries), and both solvers return identical star states $h_*= h_L = h_R$. We observe both Rusanov and HCNRS maintain the still water state up to numerical precision. In contrast, UCNRS loses the well-balanced property, with errors of about $\mathcal{O}(10^{-3})$ in both water depth and velocity. The corresponding $L^\infty$ errors are summarized in Table~\ref{tab:table2}.

Wall boundary conditions are commonly imposed using a ghost-cell approach, in which the water depth is mirrored and the velocity is reflected by a sign change. This corresponds to solving the boundary Riemann problem
\[
\mathcal{R}_{\mathrm{SWE}}(h_L = h,\; u_L = u,\; h_R = h,\; u_R = -u).
\]

For HCNRS, zero mass flux at the wall is enforced by mirror symmetry at the level of the Riemann operator. To see this, let
\[
\mathcal{R}_{\mathrm{SWE}}(h,u,h,-u) = (h_*,u_*).
\]
By mirror symmetry constraint,
\[
\mathcal{R}_{\mathrm{SWE}}(h,u,h,-u) = (h_*,-u_*),
\]
which implies \(u_* = -u_* = 0\) and therefore guarantees zero mass flux at the wall.

In contrast, UCNRS does not enforce mirror symmetry and, as a result, does not generally produce a zero intermediate velocity. This leads to a nonzero mass flux at the wall and a loss of mass conservation in the numerical scheme. This deficiency is reflected in the total mass reported in Table~\ref{tab:table2}, where UCNRS increases the total mass by $9.92\times10^{-7}$.

\begin{table}
\caption{\label{tab:table2}
Well-balanced property comparison for the still water test at $t=0.1$.
Maximum velocity and water depth errors, as well as total mass, are reported for
Rusanov, HCNRS, and UCNRS.
}
% \begin{ruledtabular}
\begin{tabular}{lccc}
 & Rusanov & HCNRS & UCNRS \\ \hline
Max $|\Delta u|$ & $1.48\times10^{-14}$ & $9.98\times10^{-15}$ & $4.68\times10^{-3}$ \\
Max $|\Delta h|$ & $1.78\times10^{-15}$ & $1.06\times10^{-14}$ & $4.42\times10^{-3}$ \\
Total mass       & $9.0$                & $9.0$                & $9.0 + 9.92\times10^{-7}$ \\
\end{tabular}
% \end{ruledtabular}
\end{table}

\begin{figure}
    \centering
    \includegraphics[width=0.8\linewidth]{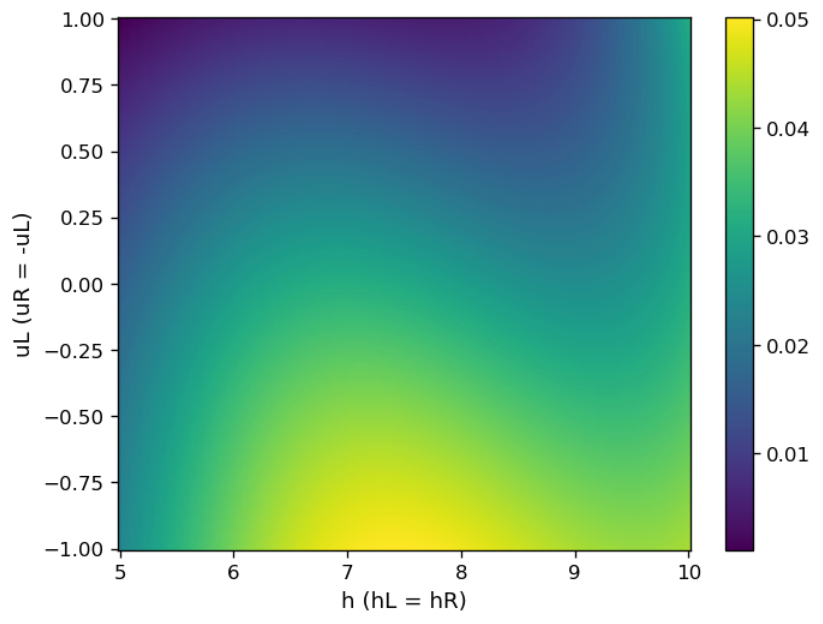}
    \caption{Absolute value of the mass flux returned by UCNRS for states with mirrored velocities,
    with $h\in[5,10]$, $u_L\in[-1,1]$, and $u_R=-u_L$.
    The nonzero flux indicates that UCNRS does not enforce zero mass flux at the wall.}
    \label{fig:flux}
\end{figure}

To further illustrate this behavior, we evaluate the mass flux returned by UCNRS for states
with mirrored velocities, where the water depth varies over $h\in[5,10]$, the left velocity satisfies $u_L\in[-1,1]$, and the right velocity is set to $u_R=-u_L$. Figure~\ref{fig:flux} shows the absolute value of the resulting mass flux. The error is not symmetric with respect to the velocity, and the flux reaches values on the order of $\mathcal{O}(10^{-2})$ with a maximum value of 0.0502.

\subsection{Radial Dam Break}

We consider the classical two-dimensional radial dam break problem and the two-dimensional shallow water equations
\begin{equation}
\partial_t \mathbf{U}
+ \partial_x \mathbf{F}(\mathbf{U})
+ \partial_y \mathbf{G}(\mathbf{U})
= 0,
\end{equation}
where
\begin{equation}
\mathbf{U} =
\begin{pmatrix}
h \\
hu \\
hv
\end{pmatrix},
\quad
\mathbf{F}(\mathbf{U}) =
\begin{pmatrix}
hu \\
hu^2 + \tfrac12 g h^2 \\
huv
\end{pmatrix},
\quad
\mathbf{G}(\mathbf{U}) =
\begin{pmatrix}
hv \\
huv \\
hv^2 + \tfrac12 g h^2
\end{pmatrix}.
\end{equation}

Here, $h$ denotes the water depth, $u$ and $v$ are the velocity components in the $x$- and $y$-directions, respectively, and $g = 9.81$ is the gravitational acceleration.

The initial condition consists of a circular region of radius $r_0 = 25\,\mathrm{m}$ centered at the origin with water depth $h_{\text{in}} = 5$, surrounded by an outer region with $h_{\text{out}} = 1$. The initial velocity is zero everywhere. Transmissive boundary conditions are applied on all sides.

To evaluate the performance of the proposed RS in a higher order discretization framework, all simulations use a second order MUSCL-Hancock scheme \cite{toro2013riemann, van1984relation} on a uniform Cartesian grid. In this predictor–corrector type framework, linear reconstruction is first used to evolve the states by half a time step, after which the RS is invoked to compute fluxes using the predicted states. The Riemann problem is solved in the normal direction, while the tangential velocity component is treated in an upwind manner. The scheme is L¹-stable under a suitable CFL and slope-limiting condition \cite{berthon2006muscl}.  Spatial reconstruction uses the monotonized central (MC) limiter. Time integration uses a CFL-based time step with $\mathrm{CFL}=0.4$. The exact RS, HCNRS, and UCNRS require 125 time steps to reach the final time $t=5$, while Rusanov requires 124 time steps.

\begin{figure}
    \centering
    \includegraphics[width=1\linewidth]{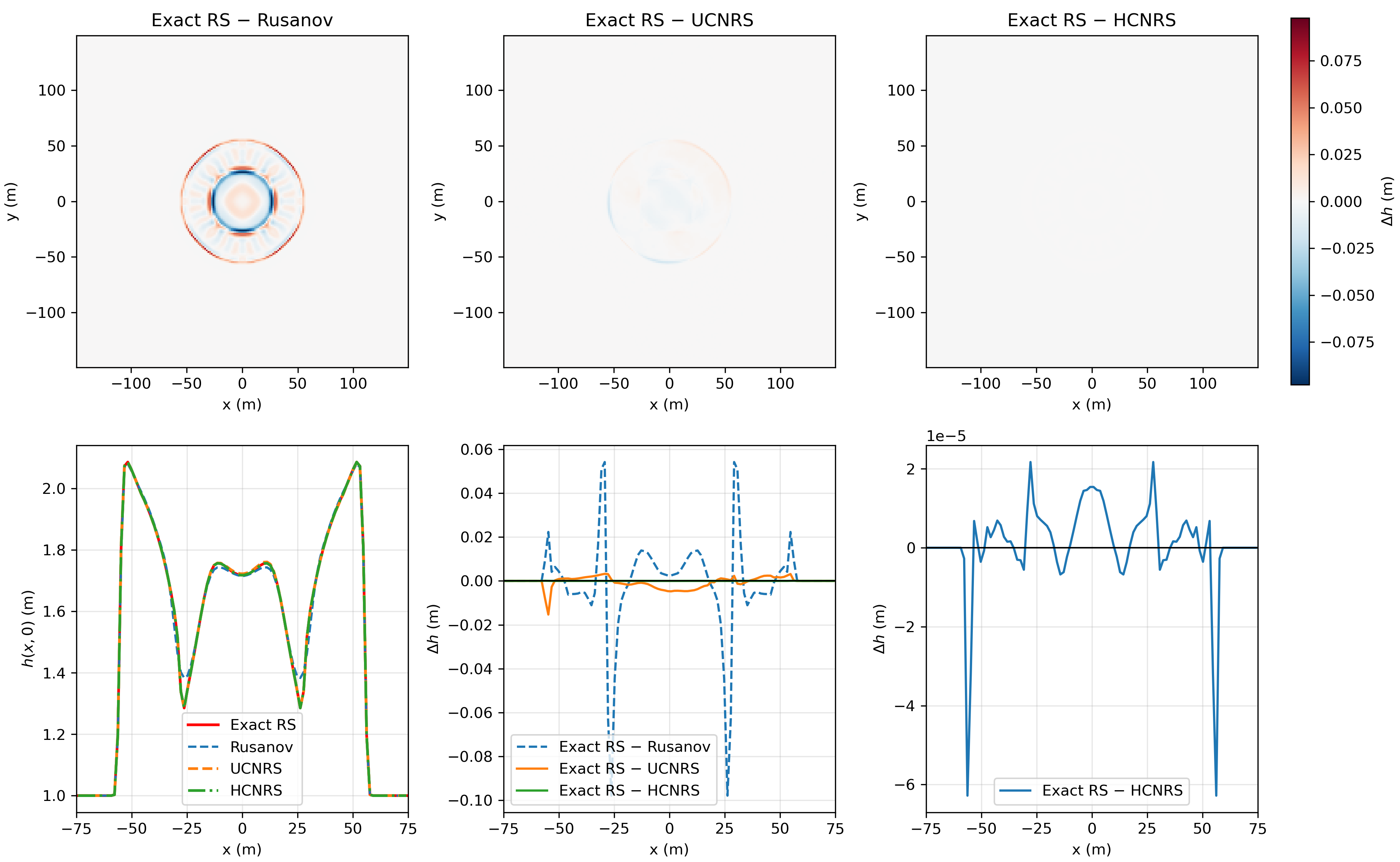}
    \caption{
    Two-dimensional dam break test at $t = 5$.
    Top row: spatial error in water depth,
    $\Delta h = h_{\mathrm{Exact RS}} - h$,
    for the Rusanov, UCNRS, and HCNRS.
    Bottom-left: midline transect of the solution, $h(x, y = 0)$.
    Bottom-middle: corresponding transect errors $\Delta h(x,0)$ for all methods.
    Bottom-right: transect error for HCNRS only.
    }
    \label{fig:dam-break}
\end{figure}

Figure~\ref{fig:dam-break} compares the numerical solutions at $t = 5$. In the two-dimensional error field, the Rusanov scheme exhibits noticeable diffusion near the expanding shock front, resulting in a smeared wave profile and a broad error band. The UCNRS substantially reduces this diffusion and produces a sharper wave structure, although small discrepancies remain in the vicinity of the wave front. In contrast, the HCNRS solution is visually indistinguishable from the exact solver at this resolution. The midline transect error reveals residual deviations for all approximate solvers. Notably, the HCNRS $L^\infty$ error is on the order of $\mathcal{O}(10^{-5})$, and Rusanov and UCNRS is on the order of $\mathcal{O}(10^{-2})$. The HCNRS error stays symmetric around $x=0$. As a result, the numerical solution maintains the radial symmetry of the problem. In contrast, UCNRS clearly shows asymmetry in the error distribution.

\subsection{Implosion}

We consider the two-dimensional compressible Euler equations
\begin{equation}
\partial_t \mathbf{U}
+ \partial_x \mathbf{F}(\mathbf{U})
+ \partial_y \mathbf{G}(\mathbf{U})
= 0,
\end{equation}
where the conserved variables and fluxes are given by
\begin{equation}
\mathbf{U} =
\begin{pmatrix}
\rho \\
\rho u \\
\rho v \\
E
\end{pmatrix},
\quad
\mathbf{F}(\mathbf{U}) =
\begin{pmatrix}
\rho u \\
\rho u^2 + p \\
\rho u v \\
u(E+p)
\end{pmatrix},
\quad
\mathbf{G}(\mathbf{U}) =
\begin{pmatrix}
\rho v \\
\rho u v \\
\rho v^2 + p \\
v(E+p)
\end{pmatrix},
\end{equation}
with the ideal-gas equation of state
\[
p = (\gamma - 1)\!\left(E - \tfrac12 \rho (u^2 + v^2)\right)
\]

Here, $\rho$ is the density, $u$ and $v$ are the velocity components in the $x$- and $y$-directions, respectively, $E$ is the total energy, $p$ is the pressure, and $\gamma$ is the ratio of specific heats.

The implosion problem was presented by Liska and Wendroff \cite{liska2003comparison}
as a benchmark for comparing eight numerical schemes. The test is demanding because it involves strong shock interactions, symmetry-driven wave focusing, and the formation of a thin jet structure.
The computational domain is the square $0 \le x \le 0.3$ and $0 \le y \le 0.3$
with solid wall boundary conditions imposed on all four sides. The gas constant is $\gamma = 1.4$

The initial condition is defined by a diagonal interface separating two constant states:
\begin{equation*}
(\rho,u,v,p)(x,y,t = 0) =
\begin{cases}
(1.0,\, 0,\, 0,\, 1.0), & \text{if } x + y > 0.15, \\[6pt]
(0.125,\, 0,\, 0,\, 0.14), & \text{if } x + y \le 0.15.
\end{cases}
\end{equation*}

To ensure consistency with the original study, simulations are conducted
on $400 \times 400$ cells.
We use the same numerical approach as in the radial dam break case. The equations are discretized
using the second-order MUSCL--Hancock finite-volume method and the MC limiter is employed. The Courant number is set to $\mathrm{CFL}=0.4$. The exact RS and HCNRS require 16,270 time steps to reach $T=2.5$, while UCNRS requires 16,100 time steps and the Rusanov requires 15,530.

Since UCNRS does not satisfy mirror symmetry, the diagonal symmetry of the implosion solution depends on the orientation of the reference frame. In the standard setup, this lack of constraint is not exposed, because the left and right states passed to the RS at each $x$-interface coincide with those at the corresponding mirrored $y$-interface (under the reflection $y=x$). To further examine this symmetry property, an additional scenario is analyzed in which the initial condition is rotated by $90^\circ$ counterclockwise, resulting in the low-density and low-pressure region being positioned in the bottom-right corner rather than the bottom-left. This configuration is denoted by $\mathrm{UCNRS(rotate)}$. In this rotated frame, the lack of mirror symmetry is exposed, and the diagonal symmetry is no longer preserved. $\mathrm{UCNRS(rotate)}$ requires 16,593 time steps to reach $T=2.5$.

\begin{figure}
    \centering
    \includegraphics[width=1.00\linewidth]{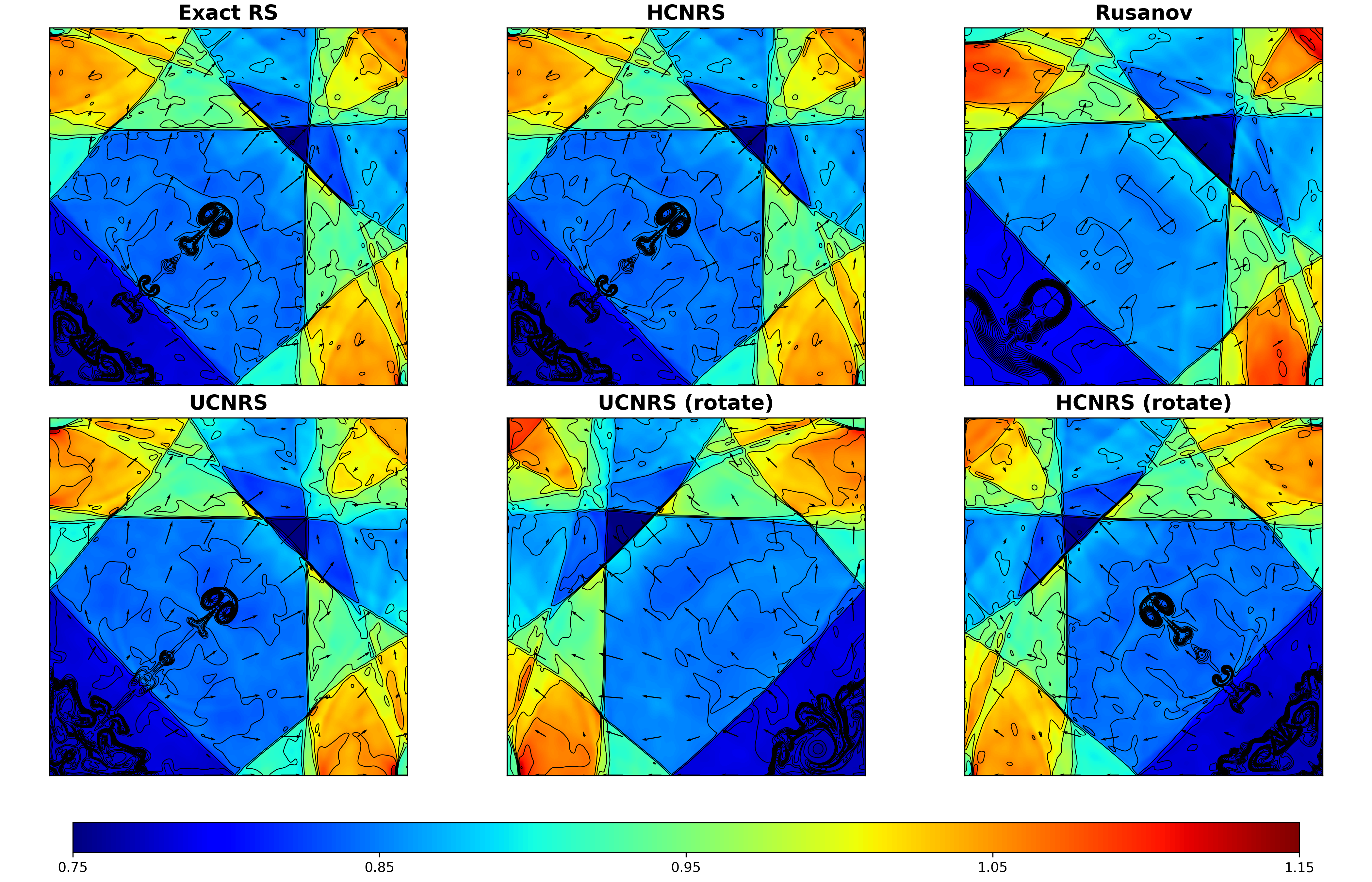}
    \caption{
    Euler implosion at $T=2.5$.
    Pressure is shown as a color map, overlaid with 31 density contours
    ($0.35$--$1.1$, step $0.025$) and velocity vectors. Formatted for comparison with Fig.~4.11 of Liska and Wendroff.
    Top row: Exact RS, HCNRS, Rusanov.
    Bottom row: UCNRS, UCNRS(rotate), HCNRS(rotate).
    Here "rotate" denotes the solution obtained from a $90^\circ$ counterclockwise rotation of the initial condition.
    }
    \label{fig:implosion-snap}
\end{figure}

Figure~\ref{fig:implosion-snap} is formatted for direct comparison with Fig.~4.11 of Liska and Wendroff, with pressure shown as a color map overlaid by density contours and velocity vectors at $T=2.5$. The exact Riemann solver with MUSCL-Hancock produces the thin diagonal jet. As documented in Fig.~4.11 of Liska and Wendroff, only two of the eight numerical schemes considered (WENO and CLAW) were able to resolve the thin jet structure. HCNRS accurately reproduces this structure, while the Rusanov scheme exhibits excessive numerical dissipation and fails to resolve the jet, highlighting the critical role of the RS. The UCNRS shifts the diagonal jet and produces a very different density structure in the bottom-left region. For UCNRS(rotate), the diagonal
symmetry is broken and the jet structure is entirely lost. In contrast, $\mathrm{HCNRS(rotate)}$ produces a solution identical to the unrotated case, just rotated by $90^\circ$, as expected. The identical nature of the exact RS and Rusanov fluxes under rotation is well-established, and thus they are omitted here for brevity.

\begin{figure}
    \centering
    \includegraphics[width=1\linewidth]{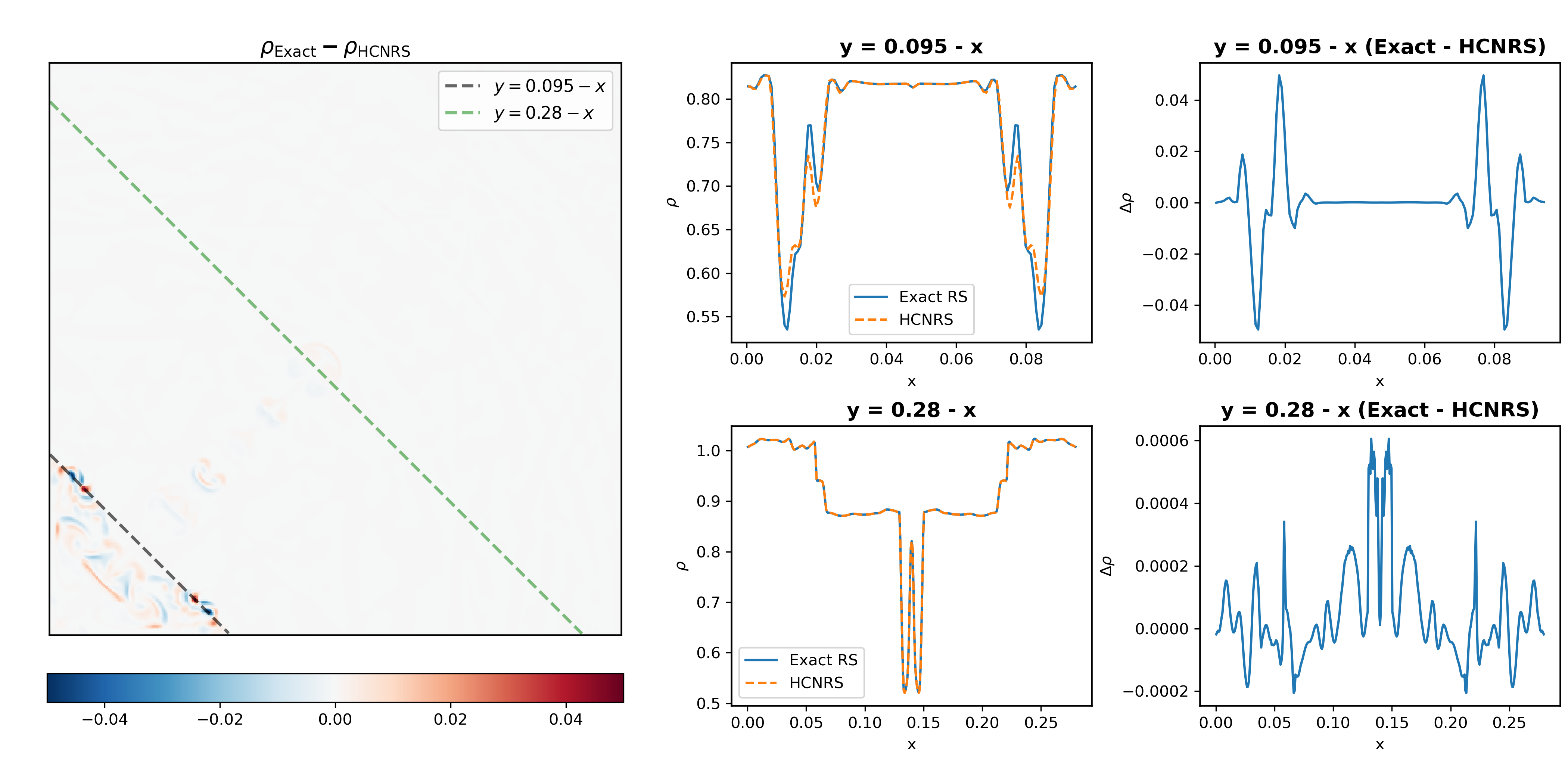}
    \caption{
    Density difference $\rho_{\mathrm{ExactRS}} - \rho_{\mathrm{HCNRS}}$
    at $T=2.5$.
    The left panel shows the two-dimensional error distribution with the two diagonal transects indicated, while the right panels display diagonal transects along
    $y=0.28-x$ and $y=0.095-x$ together with their corresponding error profiles.
    }
    \label{fig:implosion-error}
\end{figure}

To quantify the accuracy of HCNRS, we examine the density error
at $T=2.5$ by computing the difference between the exact RS
and HCNRS solutions.
Figure~\ref{fig:implosion-error} shows the spatial distribution of
$\rho_{\mathrm{Exact}} - \rho_{\mathrm{HCNRS}}$, together with two
diagonal transect plots and their corresponding error profiles.

The error is concentrated primarily in the lower-left region, where complex wave interactions occur. In contrast, the central jet structure is captured with high accuracy. The diagonal transect $y = 0.28 - x$, which passes through the core of the jet, shows excellent agreement between HCNRS and the exact RS with $L^\infty$ error is on the order of $\mathcal{O}(10^{-4})$.
The transect $y = 0.095 - x$, which traverses the region of strongest wave interaction, exhibits the largest discrepancies. HCNRS shows slightly reduced sharpness at this transect, and $L^\infty$ error is on the order of $\mathcal{O}(10^{-2})$. Nonetheless, the HCNRS shows impressive accuracy. Even after 16,270 timesteps, the model keeps high global fidelity, preserving the key structures and symmetries.

\section{Computational Cost Considerations}

We examine the computational performance of our implementations. We benchmark the evaluation time of the exact RS, Rusanov, HCNRS and UCNRS using JAX with just-in-time (JIT) compilation on a Texas Advanced Computing Center (TACC) Vista GH200 node. Random input states are sampled uniformly within admissible ranges. The average time per flux calculations from different methods are shown in Table~\ref{tab:table3}. The results show that the exact RS has the highest computational cost, with an average evaluation time of approximately 200 $\mu s$ for the shallow water and Euler equations. For the shallow water equations, both neural solvers require roughly half the evaluation time of the exact solver, while for the Euler equations, the cost is reduced to approximately two-thirds of the exact solver. The HCNRS is slightly slower than UCNRS. The Rusanov flux has the lowest evaluation time among all approaches. \cite{gyrya2024machine} reported that for the Euler equations, the exact RS averages about 340 $\mu s$ per solve and a neural network with three hidden layers (64–32–16) averages 1.48 ms on an Intel Xeon E5-2687W CPU. In our GPU-based implementation, both the exact RS and HCNRS achieve lower evaluation times, with the neural solver showing a larger speedup and achieving faster speeds than the exact RS. This difference is expected, as GPU architectures are better suited for the matrix operations involved in neural network inference. This observation is also consistent with the findings of RimNet \cite{wu5295012rimnet}, where a 15.20\%–29.66\% speedup was reported relative to an HLL solver. The efficiency gains were attributed to matrix-based neural network inference, which avoids the computational overhead of conventional RS.

\begin{table}
\caption{\label{tab:table3}
Average evaluation time per flux calculation for the shallow water and Euler equations using different Riemann solvers. Results are reported for the exact Riemann solver, the Rusanov flux, HCNRS, and UCNRS.
}
\centering
\begin{tabular}{lcccc}
 & Exact RS & Rusanov & HCNRS & UCNRS \\ \hline
Shallow Water & 188 $\mu s$ & 86.2 $\mu s$ & 109 $\mu s$ & 107 $\mu s$ \\
Euler & 209 $\mu s$ & 91.7 $\mu s$ & 155 $\mu s$ & 148 $\mu s$ \\

\end{tabular}
% \end{ruledtabular}
\end{table}   

While the observed speedup is moderate for the shallow water and ideal-gas Euler equations, this is expected because exact RS are very efficient for these systems. For many more complex systems, the exact solution incurs substantially greater computational cost, and the computational benefit of neural RS becomes considerably more pronounced. For example, in non-ideal gas dynamics and general, convex equations of state, the exact RS becomes significantly more involved. In particular, handling rarefaction waves may need the numerical integration of ordinary differential equations \cite{kamm2015exact}. As a result, the computational cost can increase substantially. For instance, \cite{ruggeri2022neural} reported benchmarks for calorically imperfect gases, supercritical fluids, and high explosives, with the latter averaging 21.36 seconds per Riemann solve. Another example occurs in multiscale methods, where the local interface phase dynamics are modeled as a Riemann problem, leading to evaluation times on the order of tens of seconds per solve \cite{magiera2016particle}.

Another potential advantage of neural RS arises in high-order schemes that employ high-order RS, such as ADER \cite{titarev2002ader}. In these methods, the computational cost of the RS increases substantially. For example, in the Toro–Titarev approach \cite{toro2002solution} to generalized Riemann problems of order $m$ (GRP$m$), a time power series solution is constructed. The leading-order term corresponds to the solution of the original Riemann problem. Higher-order terms are derived using the Cauchy–Kowalewski procedure, and each term involves solving a standard Riemann problem. This results in $m$ Riemann solves. Although approximating a GRP$m$ may require a more expressive network, the evaluation cost scales only with the network architecture. Provided that an appropriate neural network architecture is employed, the computational complexity may increase much more mildly than that of traditional GRP$m$ solvers.

\section{Conclusions}

In this work, we introduced five constraints for Riemann problems of the shallow water and ideal-gas Euler equations: positivity, consistency, mirror symmetry, Galilean invariance, and scaling invariance. Based on these constraints, we developed a neural Riemann solver that enforces these constraints strongly.

Through several benchmark problems, we demonstrated the importance of these constraints. In the still water test case, the absence of consistency leads to a loss of the well-balanced property, while the lack of mirror symmetry results in incorrect wall boundary fluxes. In the radial dam-break problem, we showed the necessity of mirror symmetry constraint for preserving radial symmetry. In the Euler implosion test case, we highlighted the critical role of an accurate RS in resolving fine-scale structures. In all cases, the proposed HCNRS, when integrated into the numerical schemes, reproduces the exact RS solutions with high accuracy. These results suggest that the neural-network-based RS provides a potential alternative to classical approaches. 

Nevertheless, limitations remain. Neural network accuracy typically degrades significantly for states lying outside the range represented in the training data. In the context of CFD, such extrapolation errors may accumulate over time and lead to simulation inaccuracies or even failure. Consequently, it is important to ensure that the admissible range of the input states is adequately covered during training, such as through prior estimation or adaptive data generation strategies. Furthermore, the theoretical stability and convergence properties of neural network based Riemann solvers within numerical schemes remain open questions.

Ongoing work focuses on extending the hard constrained framework to high order RS. Additional directions for future research include applications to more complex systems such as ideal magnetohydrodynamics, as well as the development of theoretical results on the convergence and stability of neural RS.

\section*{CRediT Authorship Contribution Statement}

\textbf{Yucheng Zhang:} Conceptualization, Methodology, Software, Validation, Formal analysis, Writing – original draft, Writing – review \& editing, Visualization. 
\textbf{Chayanon Wichitrnithed:} Methodology, Software, Validation, Writing – review \& editing. 
\textbf{Shukai Cai:} Methodology, Writing – review \& editing, Visualization. 
\textbf{Sourav Dutta:} Methodology, Writing – review \& editing.
\textbf{Kyle Mandli:} Methodology, Writing – review \& editing.
\textbf{Clint Dawson:} Methodology, Writing – review \& editing, Supervision, Project administration, Funding acquisition.

\section*{Declaration of Competing Interest}
The authors declare that they have no known competing financial interests or personal relationships that could have appeared to influence the work reported in this paper.

\section*{Acknowledgments}
This work was supported by the U.S. National Science Foundation under Award No. 2208461.

This material is based upon work supported by the NSF National Center for Atmospheric Research, which is a major facility sponsored by the U.S. National Science Foundation under Cooperative Agreement No. 1852977, and by the National Science Foundation under Grant No. 2103843 through the Rising Voices, Changing Coasts: The National Indigenous and Earth Sciences Convergence Hub.

The authors acknowledge the Texas Advanced Computing Center (TACC) at The University of Texas at Austin for providing computational resources that have contributed to the research results reported within this paper. URL: http://www.tacc.utexas.edu

\section*{Data Availability}
Data will be made available on request.

\appendix

\section{Rusanov Flux}
\label{app:rusanov}

The Rusanov flux, also known as the local Lax--Friedrichs flux, is widely used as an approximate Riemann solver in CFD simulations and serves as a baseline in this work. This solver accounts only the fastest wave speed between left and right waves. Given left and right states $\mathbf{U}_L$ and $\mathbf{U}_R$, the flux is defined as
\begin{equation}
\hat{\mathbf{F}}_{\mathrm{Rus}}(\mathbf{U}_L,\mathbf{U}_R)
=
\frac{1}{2}\left(
\mathbf{F}(\mathbf{U}_L) + \mathbf{F}(\mathbf{U}_R)
\right)
-
\frac{1}{2}\alpha
\left(
\mathbf{U}_R - \mathbf{U}_L
\right),
\end{equation}
where $\alpha$ is the maximum wave speed and $\mathbf{F}(\mathbf{U})$ is the corresponding flux function defined in Eq.~\eqref{eq:conservation_law}.

The wave speeds are given by the eigenvalues of the flux Jacobian. For the shallow water equations,
\begin{equation}
\alpha =
\max\left(
|u_L| + \sqrt{g h_L},
|u_R| + \sqrt{g h_R}
\right).
\end{equation}

For the Euler equations,
\begin{equation}
\alpha =
\max\left(
|u_L| + c_L,
|u_R| + c_R
\right),
\end{equation}
where \begin{equation}
c_L = \sqrt{\frac{\gamma p_L}{\rho_L}}, \qquad
c_R = \sqrt{\frac{\gamma p_R}{\rho_R}}
\end{equation}
are the sound speeds corresponding to the left and right states.

\bibliographystyle{elsarticle-num}
\bibliography{references}


%% For numbered reference style

\end{document}